\documentclass[fleqn,10pt]{wlscirep}
\usepackage[utf8]{inputenc}
\usepackage[T1]{fontenc}
\usepackage{amssymb,amsmath,amsbsy,amsgen,amsfonts,amsthm,mathtools}

\newcommand{\abs}[1]{\left\vert #1\right\vert}
\newcommand{\bra}[1]{\left\langle{#1}\right\vert}
\newcommand{\ket}[1]{\left\vert{#1}\right\rangle}

\newcommand{\mean}[1]{\langle#1\rangle}

\title{Theoretical studies on quantum imaging with time-integrated single-photon detection under realistic experimental conditions}

\author[1,*]{Byeong-Yoon Go}
\author[2,$\dagger$]{Changhyoup Lee}
\author[1,$\ddagger$]{Kwang-Geol Lee}
\affil[1]{Department of Physics, Hanyang University, Seoul 04763, Republic of Korea}
\affil[2]{Korea Research Institute of Standards and Science, Daejeon 34113, Republic of Korea}
\affil[*]{Present address: Korea Advanced Institute of Science and Technology, Daejeon 34141, Republic of Korea}
\affil[$\dagger$]{changhyoup.lee@gmail.com}
\affil[$\ddagger$]{kglee@hanyang.ac.kr}

\begin{abstract}
We study a quantum-enhanced differential measurement scheme that uses quantum probes and single-photon detectors to measure a minute defect in the absorption parameter of an analyte under investigation. For the purpose, we consider two typical non-classical states of light as a probe, a twin-Fock state and a two-mode squeezed vacuum state. Their signal-to-noise ratios (SNRs) that quantifies the capability of detecting the defect are compared with a corresponding classical imaging scheme that employs a coherent state input. A quantitative comparison is made in terms of typical system imperfections such as photon loss and background noise that are common in practice. It is shown that a quantum enhancement in SNR can be described generally by the Mandel Q-parameter and the noise-reduction-factor, which characterize an input state that is incident to the analyte. We thereby identify the conditions under which the quantum enhancement remains and can be further increased. We expect our study to provide a guideline for improving the SNR in quantum imaging experiments employing a differential measurement scheme with time-integrated single-photon detectors.
\end{abstract}
\begin{document}

\flushbottom
\maketitle
%
%
\thispagestyle{empty}


\section*{Introduction}

The capability to measure a small optical signal is of utmost importance in both fundamental studies and practical applications. For example, a microscopy technique measuring optical responses of an object has led to a wide range of development in the fields of biochemical and medical sciences~\cite{Mockl2014,Min2011,Duan2018,Wu2019}. The highly sensitive optical interferometer also enables the detection of gravitational waves~\cite{Abbott2016}. However, the signal quality obtainable with a classical probe is fundamentally limited by the shot noise even when all technical noises are removed. The shot noise is originated from the discretized nature of light that composes a classical probe, i.e., a coherent state whose photon number statistics follows a Poisson distribution. The associated signal-to-noise ratio (SNR) is proportional to the square root of the intensity,~$\sqrt{I}$, so that the SNR can be enhanced by increasing the intensity of light used for the measurement. However, increasing the optical power is not always acceptable due to optical damages or unwanted photo-sensitive effects that can be caused by the high intensity at incidence~\cite{Liu1995,Taylor2016,Braun2008}. In such limited situations, an alternative way to reach an aimed SNR instead of increasing the optical intensity has been suggested~\cite{Caves1980,Caves1981} and it is to use quantum probes with exploiting useful quantum properties of light, which are absent in classical probes~\cite{Lee2021}.

A number of experimental demonstrations for quantum sensing and imaging have shown that probing photo-sensitive samples with quantum light can achieve sub-shot-noise limited behaviors in various aspects such as sensitivity or precision~\cite{Lee2021}. The most widely used quantum probe is the so-called twin-beam that possesses the complete photon-number correlation between a signal and an idler mode~\cite{Meda2017}. They have been used in quantum imaging with an intensity-sensitive measurement~\cite{Lounis2005, Polyakov2009, Chu2017, Lee2017, Lee2018, Lawrie2019} and in quantum enhanced absorption measurement scheme~\cite{Tapster1991,Moreau2017,Losero2018}, while also been exploited in quantum phase estimation with phase-sensitive measurement~\cite{Hudelist2014, Kok2002, Giovannetti2004, Xiang2011}. Another useful quantum state in quantum imaging is a Fock state whose photon number statistics is sub-Poissonian. The Fock state is known to be optimal for intensity-sensitive measurement~\cite{Adesso2009}. Although the generation of an arbitrary sub-Poissonian field with high photon number has been studied theoretically and experimentally~\cite{Perina2017}, it is still demanding in practice~\cite{Varcoe2000, Bertet2002,Waks2006}. The most common quantum optical detector, on the other hand, is a single-photon threshold detector which distinguishes between vacuum ($n=0$, i.e., ‘no click’) and photons ($n \ge 1$, i.e., ‘click’)~\cite{Hadfield2009, Eisaman2011}. It has been realized by an avalanche photodiode~\cite{Bachor2004} or a superconducting nanowire single-photon detector~\cite{Natarajan2012}. When multiple single-photon detectors are combined via a multiplexing scheme, they can provide pseudo photon number statistics~\cite{Fitch2004,Achilles2004} or click statistics~\cite{Sperling2012, Lee2016}. A single-photon detector is also often used in a simpler way for time-integrated detection, which counts the number of ‘click’ events for an exposure time~$T=m \Delta t$, where~$m$ denotes the number of segment and~$\Delta t$ represents the unit of time interval. The latter scheme has been applied to various sensing and imaging applications, e.g., single-photon quantum imaging~\cite{Yang2020}.

In the ideal lossless case, the aforementioned quantum resources exhibit remarkable behaviors including sub-shot-noise limited SNR~\cite{Brida2010}, but inevitable losses that are present in reality degrade the quantum enhancement in SNR~\cite{Berchera2019, Genovese2016}. For example, a limited detection efficiency of a single-photon detector reduces a quantum gain since quantum features are very vulnerable to loss or decoherence~\cite{Dorner2009, Datta2011, Peter2011}. Therefore, the losses need to be taken into account in quantification of quantum enhancement, which may suggest a guideline for the use of quantum resources in various imaging schemes. The effects of realistic imperfections have been considered in various manners, for example, finding optimal quantum probes for estimation of a lossy phase shift~\cite{Dorner2009, Kacprowicz2010}, finding nearly optimal measurements in lossy Mach-Zehnder interferometer~\cite{Gard2017}, identifying optimal Gaussian resources in phase measurement~\cite{Oh2019, Oh2020}, improving indistinguishability of interfering photons~\cite{Jachura2016}, robust preparation of the NOON state~\cite{Ulanov2016}, and reaching the ultimate quantum limit in chirality sensors measuring, e.g., optical activity~\cite{Yoon2020} and circular dichroism~\cite{Ioannou2020}.

Along with theses, we investigate in this work the effect of losses and imperfections that are very common in practical experimental setups for quantum imaging. For the purpose, we consider a quantum imaging scheme with a twin beam differential measurement that can effectively eliminate the common excess noise~\cite{Bachor2004}. As a detection scheme, we employ time-integrated single photon detection. Such a detection scheme has been used in a recent experimental study that investigates the photon number distribution and resultant non-classical features for compound twin-beams~\cite{Perina2021}. There, the compound twin-beams are composed of identical twin-beams that are sufficiently weak, used to substitute stronger genuine twin beams that require the use of photon-number-resolving detectors. On the other hand, our work focuses on establishing a theoretical model of the time-integrated single-photon detection scheme for various quantum states to analyze the effects of noisy quantum sensing environments. In the presence of losses and imperfections, we compare the SNRs for two useful quantum states that are known to achieve quantum enhancement in intensity-sensitive measurement~\cite{Lee2017}; a two-mode squeezed vacuum (TMSV) state which can be produced by a spontaneous parametric down-conversion (SPDC) process~\cite{Lee2017, Lee2016b, Gao1998, Brambilla2008}, and a twin-Fock (TF) state which can be produced by various kinds of photon sources. One can see that both quantum states exhibit maximum photon-number correlation between the two modes, which is very useful to minimize the uncertainty in differential measurement. The reduced uncertainty consequently enhances the SNR, which we characterize in this work by two parameters; Mandel Q-parameter~\cite{Mandel1979} and the noise reduction factor~$\sigma$~\cite{Jedrkiewicz2004}, which have been ubiquitously used to distinguish various states of light. We then study the effect of loss and background noise (here, dark count) in the SNR when probing a sample with the two quantum states. Furthermore, we identify when a quantum enhancement remains in comparison with a classical imaging scheme that employs a coherent state of light.

\section*{Theoretical Model}

\subsection*{Imaging Setup}
\begin{figure}[t]
\centering
\includegraphics[width=0.45\textwidth]{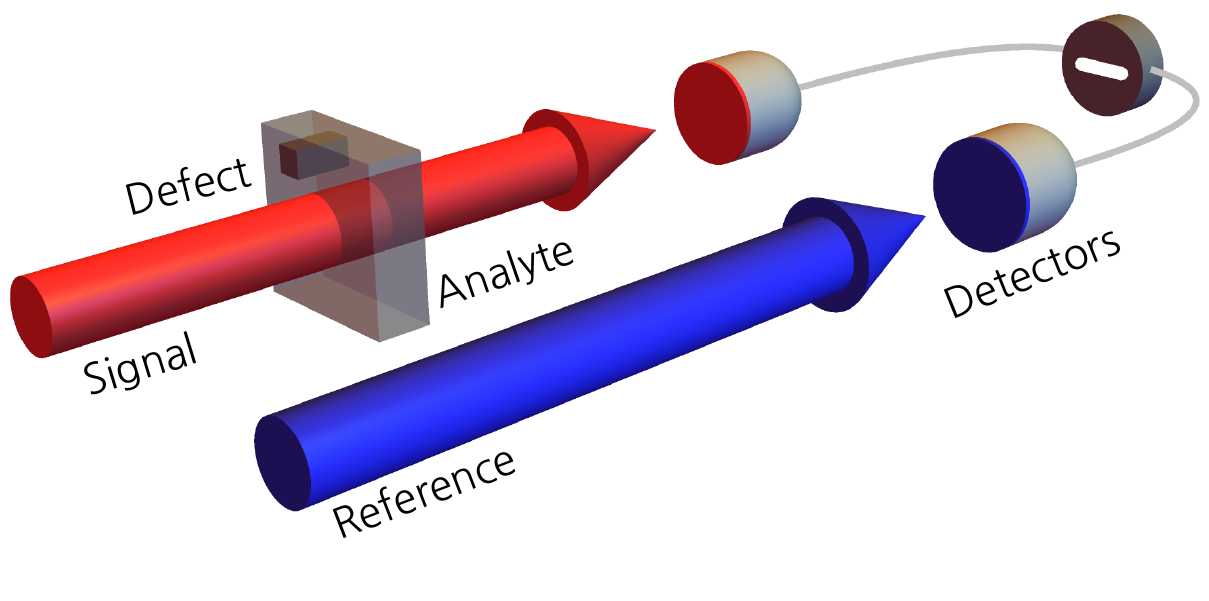}
\caption{
Differential detection scheme for imaging of an absorptive analyte with minute defects. The signal (red) beam passes through the analyte, whereas the reference (blue) beam is kept unchanged. The intensities of the transmitted beams are measured at detectors, yielding their difference via post-processing. 
The mathematical formulation of the considered imaging scheme is described in the main text, where the photon loss and the background noise are also properly considered. 
}
\label{fig:fig1}
\end{figure}

We consider an intensity-difference measurement scheme to detect a minute defect~$\delta\alpha$ of the absorption parameter~$\alpha$ of an absorptive sample, as depicted in Fig.~\ref{fig:fig1}. For simplicity, we assume that the two beams are collimated, so that they can be treated respectively as one-dimensional photon fluxes. Let us now consider an arbitrary two-mode state that consists of a signal (label `1') and a reference (label `2') mode, written in the basis of two-mode Fock states as
\begin{align*}
\ket{\Psi}=\sum_{n_1,n_2=0}^{\infty}C_{n_1,n_2}\ket{n_1}_1\ket{n_2}_2,
\end{align*}
where~$\sum_{n_1,n_2=0}^{\infty}\abs{C_{n_1,n_2}}^2=1$. Here, the mean photon number~$N_j$ of~$j$th mode is given as
\begin{align*}
N_j=\bra{\Psi}\hat{n}_j\ket{\Psi}_\text{in},
\end{align*}
where~$\hat{n}_j=\hat{a}_j^\dagger\hat{a}_j$ is the photon number operator and~$\hat{a}_j$ is the annihilation operator satisfying the bosonic commutation relation~$[\hat{a}_j,\hat{a}_k^\dagger]=\delta_{jk}$ for~$j,k\in\{1,2\}$. The signal mode of the two-mode state passes through an absorptive sample whose transmission coefficient is~$t_1$, while the idler mode is kept unchanged as a reference, as depicted in Fig.~\ref{fig:fig1}. The interaction with the absorptive sample under study can be described in the Heisenberg picture by the relation between the operators~$\hat{a}_1$ and~$\hat{a}'_1$ of the modes before and after hitting the sample, respectively, written as~\cite{Meda2017},
\begin{align*}
\hat{a}'_1=t_1\hat{a}_1+i\sqrt{1-t_1^2}\hat{v}_1,
\end{align*}
where~$\hat{v}_1$ denotes an operator for a virtual mode associated with absorption. Such a relation implies that the absorption occurs for individual photons probabilistically with a rate~$\alpha=1-t_1^2$, finally modifying the photon number distribution of the transmitted field. The lossless idler mode is described by an operator kept unchanged, i.e.,~$\hat{a}'_2=\hat{a}_2$, and is used as a reference to measure the change of intensity on the signal mode. Therefore, the intensity-difference~$I_{-}$ to be measured is described by
\begin{align*}
I_{-}=\mean{\hat{n}'_-}_\text{out}=\mean{\hat{n}'_2}_\text{out}-\mean{\hat{n}'_1}_\text{out},
\end{align*}
and its noise is given as
\begin{align*}
(\Delta I_{-})^2 
&=\langle (\Delta \hat{n}'_-)^2\rangle_\text{out}
=\langle (\Delta \hat{n}'_1)^2\rangle_\text{out}+\langle (\Delta \hat{n}'_2)^2\rangle_\text{out}-2\text{Cov}(\hat{n}'_1,\hat{n}'_2)_\text{out},
\end{align*}
where~$\text{Cov}(X,Y)=\mean{XY}-\mean{X}\mean{Y}$ denotes the covariance between~$X$ and~$Y$, and the subindex `in (out)' represents the expectation value calculated with respect to an input (output) state. The individual terms can be calculated as
\begin{align*}
\langle (\Delta \hat{n}'_1)^2\rangle_\text{out}&=(1-\alpha)^2\langle (\Delta \hat{n}_1)^2\rangle_\text{in}+\alpha(1-\alpha)\mean{\hat{n}_1}_\text{in},\\
\langle (\Delta \hat{n}'_2)^2\rangle_\text{out}&=\langle (\Delta \hat{n}_2)^2\rangle_\text{in},\\
\text{Cov}(\hat{n}'_1,\hat{n}'_2)_\text{out}&=(1-\alpha)\text{Cov}(\hat{n}_1,\hat{n}_2)_\text{in}.
\end{align*}
From now on, we particularly concentrate on twin-mode states, which have the equal mean photon number and equal photon number variance in both modes, i.e., $\mean{\hat{n}_1}_\text{in}=\mean{\hat{n}_2}_\text{in}=N$ and~$\langle (\Delta \hat{n}_1)^2\rangle_\text{in}=\langle (\Delta \hat{n}_2)^2\rangle_\text{in}$. Such an assumption applies to most imaging scenarios, where the signal mode is compared with the reference mode. The above signal and noise can thereby be simplified as
\begin{align}
I_{-}&=\mean{\hat{n}'_2}_\text{out}-\mean{\hat{n}'_1}_\text{out}=\alpha N,\nonumber\\
(\Delta I_{-})^2 &=N\left[\alpha^2Q+2(1-\alpha)\sigma+\alpha\right],\nonumber
\end{align}
where~$Q=\langle (\Delta \hat{n})^2\rangle_\text{in}/\mean{\hat{n}_1}_\text{in}-1$ denotes the Mandel Q-parameter~\cite{Mandel1979} and~$\sigma=\langle (\Delta \hat{n}_-)^2\rangle_\text{in}/(\mean{\hat{n}_1}_\text{in}+\mean{\hat{n}_2}_\text{in})$ represents the noise reduction factor (NRF)~\cite{Jedrkiewicz2004}. 
Consequently, the SNR for~$I_{-}$ can be written as~\cite{Brambilla2008}
\begin{align}
\text{SNR}= \frac{I_{-}}{\Delta I_{-}} =\frac{\alpha\sqrt{N}}{\sqrt{\alpha^2Q+2(1-\alpha)\sigma+\alpha}}.
\label{eq:SNR}
\end{align}
Note that $Q\ge0$ for all classical lights, so the light exhibiting sub-Poissonian photon number statistics, i.e.,$-1\le Q<0$, is called non-classical. Also, the NRF is greater than unity (i.e.,~$\sigma\ge1$) for all pure classical states (i.e., a product of two arbitrary coherent states). Hence, the product of two identical coherent states can be characterized by $Q=0$ and $\sigma=1$. One can thus find useful pure quantum states of light with~$-1\le Q<0$ and ~$0\le\sigma<1$, improving the SNR of Eq.~\eqref{eq:SNR}. These characteristic parameters allow one to understand the role of quantum effects in enhancing the SNR in the differential measurement scheme for an absorption parameter~\cite{Brambilla2008,Brida2010}.

\begin{figure}[t]
\centering
\includegraphics[width=0.45\textwidth]{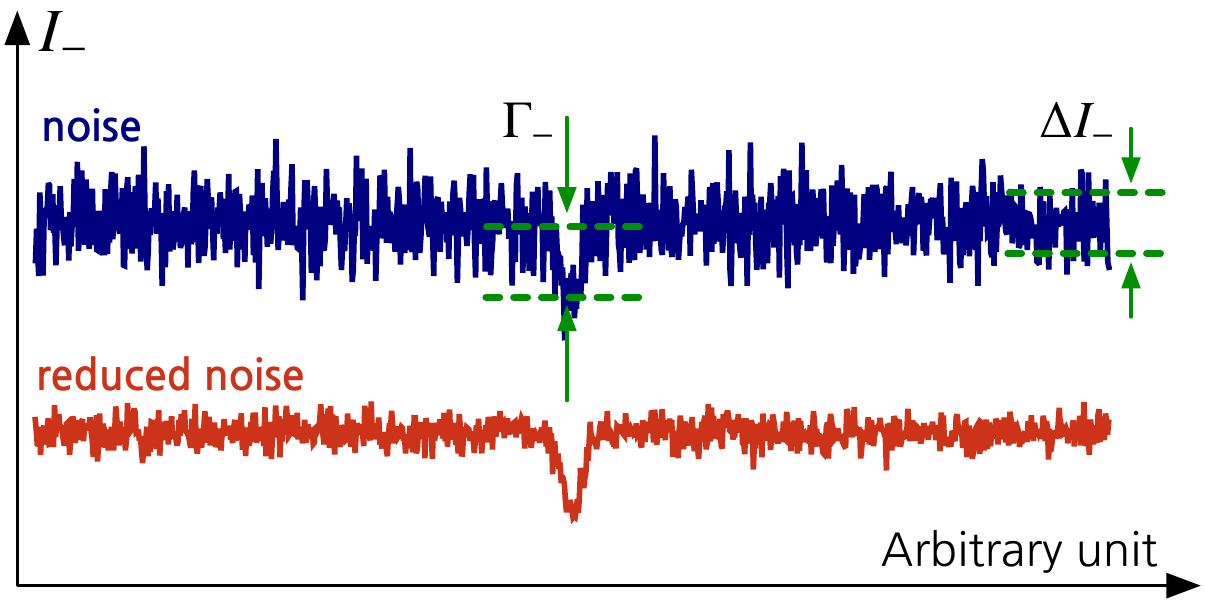}
\caption{Illustration of the measurement outcomes of the intensity-difference~$I_{-}$ that would be obtained with the noise~$\Delta I_{-}$ for a given state of light illuminated into an absorptive analyte with~$\alpha$. The presence of the defect~$\delta \alpha$ is represented by a dip with a depth of~$\Gamma_{-}$ in the noisy outcomes. If~$\Gamma_{-}$ is not large enough with respect to the noise~$\Delta I_{-}$, the presence of the defect cannot be recognized or is hardly distinguished (e.g., upper outcomes). Therefore, the use of a particular state of light yielding the reduced noise (e.g., lower outcomes) enables the defect to be more detectable under the noisy background. 
}
\label{fig:fig2}
\end{figure}

Let us now introduce a minute defect~$\delta\alpha$ in the absorption parameter~$\alpha$ of the sample under investigation as in Fig.~\ref{fig:fig1}. The defect~$\delta\alpha$ can be detected by comparing the transmitted intensities between the two positions in the samples, i.e., between ones with and without the defect. In other words, the transmitted signal having passed through the point with the defect needs to be distinguishable from the point with no defect. Their difference~$\Gamma_{-}$ reads
\begin{align*}
\Gamma_{-}=\mean{\hat{n}'_-(\alpha+\delta\alpha)}_\text{out}-\mean{\hat{n}'_-(\alpha)}_\text{out}=\delta\alpha N.
\end{align*}
It is obvious that the defect~$\delta\alpha$ is detectable when the difference~$\Gamma_{-}$ is greater than the photon number noise~$\Delta I_{-}$, as depicted in Fig.~\ref{fig:fig2}. The larger the ratio of~$\Gamma_{-}$ to~$\Delta I_{-}$ is, the more detectable the defect~$\delta \alpha$ is. Hence, we thus define the~$\text{SNR}^{*}$ to quantify the capability of detecting a minute defect as
\begin{align}
\text{SNR}^{*} = \frac{\Gamma_{-}}{\Delta I_{-}}=\frac{\delta\alpha}{\alpha}\text{SNR}.
\label{eq:SNR_star}
\end{align}
This clearly indicates that the defect in the absorptive sample is more detectable when probing with light exhibiting small~$Q$ and~$\sigma$. It is known that the smallest value of~$Q$ and~$\sigma$ can be achieved by the TF states~$\ket{N}\ket{N}$~\cite{Lee2021}. However, experimental generation of large Fock states with~$N\gg1$ is demanding within current technology~\cite{Varcoe2000, Bertet2002,Waks2006}. Alternatively, one can use~$N$ single-photons~\cite{Lounis2005,Eisaman2011,Meyer-Scott2020}, or the TMSV state --- the most common quantum probe in practical quantum imaging, which can be written as
\begin{align}
\vert \text{TMSV}\rangle 
= \sum_{n=0}^{\infty}\frac{(-e^{i\theta_\text{s}}\tanh r)^n}{\cosh r}\vert n,n\rangle,\label{eq:TMSV}
\end{align}
where~$r$ denotes the squeezing parameter and~$\theta_\text{s}$ represents the squeezing angle.
The TMSV state has been widely used in many applications including quantum imaging~\cite{Genovese2016}, quantum illumination~\cite{Lloyd2008,Tan2008,Lopaeva2013,Nair2020}, and quantum sensing~\cite{Meda2017}. They can be generated from a SPDC~\cite{Burnham1970, Heidmann1987, Schumaker1985} and the characteristic parameters~$r$ and~$\theta_\text{s}$ can be tuned in a controlled manner in experiments. The enhanced SNR by the use of TMSV states comes from the strong correlation in photon number between the signal and reference mode~\cite{Jedrkiewicz2004,Bondani07,Blanchet08,Perina12}, leading to~$\sigma=0$ regardless of the squeezing strength~$r$. Despite the reduced NRF, the Mandel~Q-parameter reads~$Q=N$ with a mean photon number~$N=\sinh^2 r$ for the TMSV state of Eq.~\eqref{eq:TMSV}, i.e., large~$r$ is rather detrimental.

\begin{figure*}[b]
\centering
\includegraphics[width=0.85\textwidth]{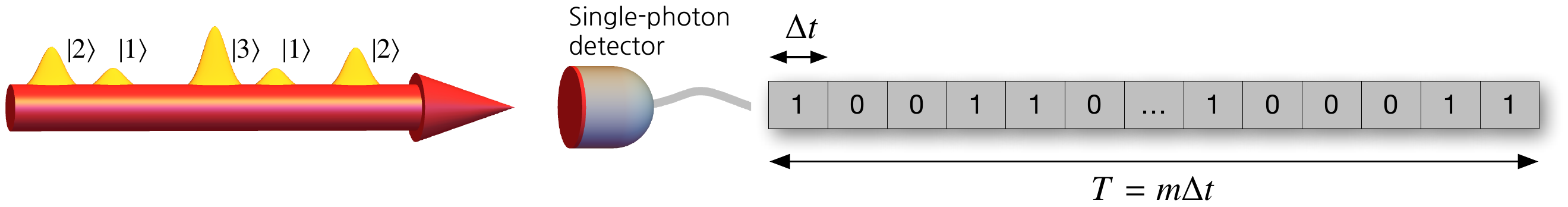}
\caption{
Time-integrated on/off detection is performed at each mode by a single photon counting module (SPCM) for an exposure time~$T$ that consists of~$m$ intervals of~$\Delta t$, i.e.,~$T=m\Delta t$. The sequential measurement outcomes are obtained for time~$T$, where the outcomes `$0$' and `$1$' denote the absence and the presence of photons, respectively.
}
\label{fig:fig3}
\end{figure*}

\subsection*{Differential measurement with time-integrated single-photon detection}

The aforementioned imaging scheme implicitly assumes the capability of resolving photon numbers in detection, represented by the photon number operator~$\hat{n}_j$. It can be achieved by superconducting transition-edge-sensors~\cite{Miller2003,Rosenberg2005,Divochiy2008,Fukuda2011, Calkins2013} with high efficiency, but they are still far from being widely used. A more commonly used detector in a number of quantum optics experiments is a single-photon threshold detector~\cite{Natarajan2012, Hadfield2009}, which can be realized by commercially available single-photon-counting module (SPCM). Therefore, it is more practical to consider the above differential measurement scheme to be made by two identical SPCMs. We note that the SPCM suffers from the dead time in the order of ~$10^2\sim10^3$~ns, for which no detection can occur and which comes right after every detection fired. With such a feature, one can consider a time-integrated detection scheme that counts the number~$c$ of `click' outcomes for an exposure time~$T= m \Delta t$, where~$m$ sequential detections with an interval~$\Delta t$ are made, as depicted in Fig.~\ref{fig:fig3}. This scheme provides the number of `click' events as a measurement outcome. In general, the distribution of~$c$ is different from the true photon number distribution obtainable by the aforementioned photon number resolving detector. They however become almost equal in the limit of~$N\ll 1$, where~$\text{SNR}$ of Eq.~\eqref{eq:SNR} and~$\text{SNR}^{*}$ of Eq.~\eqref{eq:SNR_star} can be used for the time-integrated single-photon detection scheme (see supplementary information for a rigorous justification). We therefore assume the small average photon number for all the states considered in this work, enabling to use the time-integrated detection scheme with a SPCM as an approximate photon number resolving detector. 

Let us now consider the values of~$Q$ and~$\sigma$ for the three states considered above. First, for the TF state input~$\vert N\rangle\vert N\rangle$, the SPCM always fires a detection event for any~$N\ge 1$, i.e., no difference among~$N$'s. When the TF states are consecutively injected with temporal spacing longer than the dead time of SPCMs, one can obviously see that~$Q_\text{TF}=-1$, and~$\sigma_\text{TF}=0$. Again, these results are obtainable for any~$N\ge1$, so we focus on the simplest, but the most practical case of~$N=1$, i.e.,~$\vert 1\rangle \vert 1 \rangle$. Second, for the coherent state input~$\vert \alpha \rangle \vert \alpha\rangle$,~$Q_\text{coh}=0$ originated from the Poisson photon number statistics and~$\sigma_\text{coh}=1$ due to the absence of the photon number correlation between the two modes. Third, for the TSMV state input, the reduced state for the individual mode is a thermal state, i.e.,~$\rho_1=\text{Tr}_2[\vert \text{TMSV}\rangle\langle \text{TMSV}\vert]=\rho_\text{th}$ under single-mode approximation~\cite{Avenhaus2008}. However, the down-converted photon pairs produced by the SPDC process consist of multiple (in fact, continuous) frequencies in practice, i.e., the reduced state is given by a statistically mixed thermal states with the respective spectral modes. In most cases, the temporal detection resolution of SPCMs is much longer ($\sim10^{-9}$s) than the coherence time of the down-converted photons for each mode ($\sim10^{-12}$s), so the statistical features of the individual modes are washed out. Therefore, the measurement results at individual modes are almost equal to the coherent state, i.e., we can thus put~$Q_\text{TMSV}=0$. Nevertheless, the photon number correlation is still preserved at individual spectral modes, so~$\sigma_\text{TMSV}=0$.

\begin{figure*}
\centering
\includegraphics[width=0.45\textwidth]{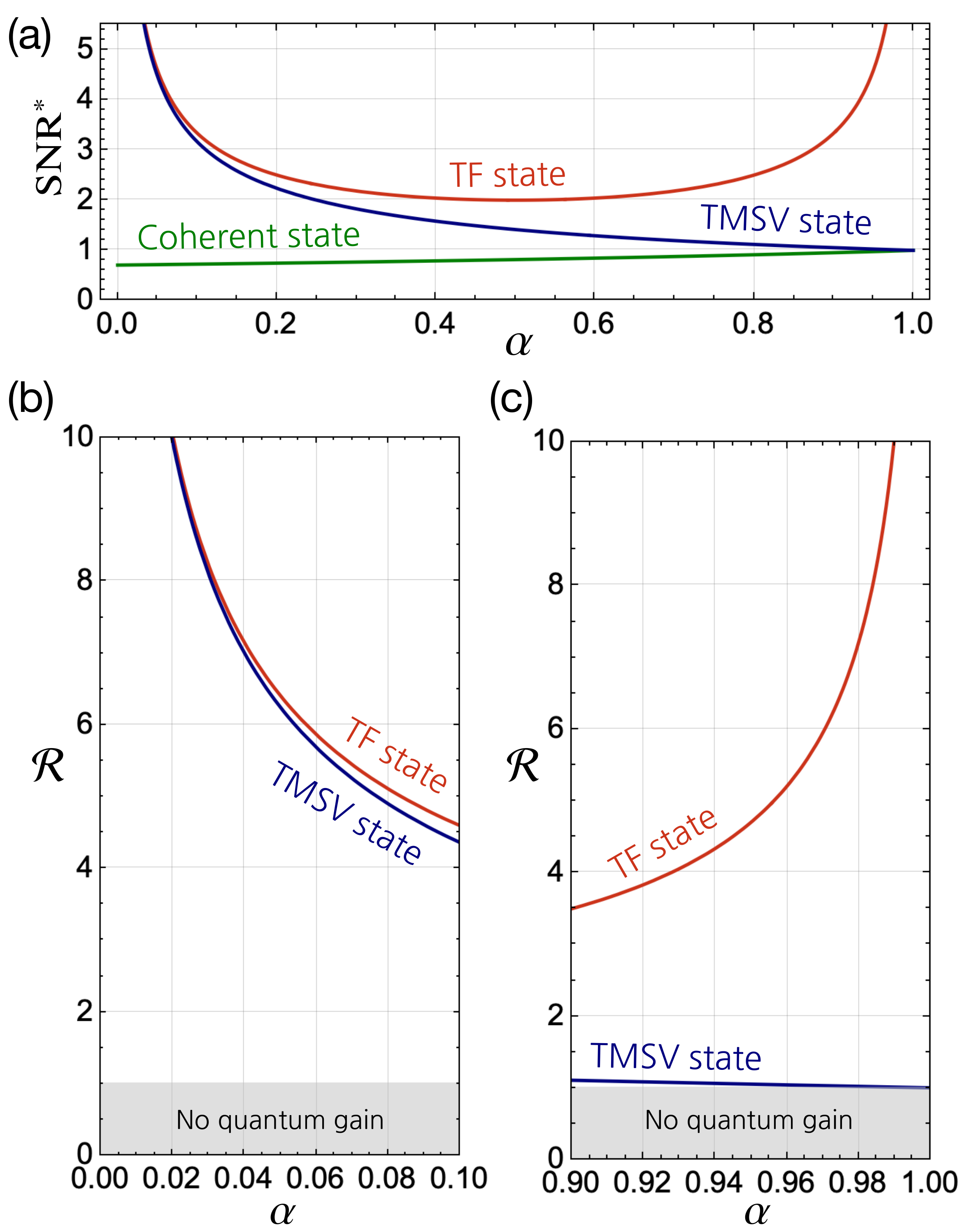}
\caption{
(a) The~$\text{SNR}^{*}$s obtainable when probing the analyze with a coherent state, the TF state, and the TMSV state as a function of the sample absorption~$\alpha$. 
The quantum gain~$\mathcal{R}$ of the~$\text{SNR}^{*}$ with respect to the benchmark~$\text{SNR}^{*}_\text{coh}$ is elaborated on two particular regimes: (b) small~$\alpha$ and (c) large~$\alpha$.
Here,~$\delta\alpha=10^{-3}$ and~$\bar{c}=10^6$ (i.e.,~$N_\text{eff}=1$) are considered as an example.
}
\label{fig:fig4}
\end{figure*}

Therefore, the~$\text{SNR}^{*}$s with respect to the three input states can be written as 
\begin{align*}
\text{SNR}^{*}_{\text{coh}}=\frac{\sqrt{N_\text{eff}}}{\sqrt{2-\alpha}},\\
\text{SNR}^{*}_{\text{TF}}=\frac{\sqrt{N_\text{eff}}}{\sqrt{\alpha-\alpha^2}},\\
\text{SNR}^{*}_{\text{TMSV}}=\frac{\sqrt{N_\text{eff}}}{\sqrt{\alpha}},
\end{align*}
where~$N_\text{eff}=(\delta\alpha)^2 \bar{c}$ denotes the effective mean photon number involved in detection for the average click number $\bar{c}$. 
The above~$\text{SNR}^{*}$s are compared in Fig.~\ref{fig:fig4}(a) in the absence of system loss and background noise. It is clear that using the considered quantum states exhibits larger~$\text{SNR}^{*}$ than the case using a coherent state input. More specifically, the~$\text{SNR}^{*}_\text{TF}$ becomes larger
as~$\alpha$ increases or decreases from~$\alpha=0.5$ and eventually diverges in the limit of small and large~$\alpha$. This is because the photon number uncertainty of the transmitted TF state becomes zero when~$\alpha$ approaches zero (i.e., highly transparent) or unity (i.e., highly absorbing). On the other hand, the~$\text{SNR}^{*}_\text{TMSV}$ monotonically decreases as the absorption increases. It is interesting to see that~$\text{SNR}^{*}_\text{TMSV}\simeq \text{SNR}^{*}_\text{TF}$ in the limit of small absorption ($\alpha\simeq 0$, i.e., highly transparent), whereas~$\text{SNR}^{*}_\text{TMSV}\simeq \text{SNR}^{*}_\text{coh}$ in the limit of large absorption ($\alpha\simeq 1$, i.e., highly absorbing). This is because when~$\alpha\simeq 0$, the transmitted TMSV state has the strong photon number correlation between the two modes, whereas when~$\alpha\simeq 1$, the correlation is almost destroyed and the detected photon statistics are almost the same as the transmitted coherent state. Although it is evident that the case using the TF state outperforms the cases using the other states at any value of~$\alpha$, the TMSV state can alternatively be used instead of the TF state when the absorption is small, i.e.,~$\alpha\ll1$.

The amount of quantum gain in~$\text{SNR}^{*}$ obtainable using a twin-mode state can be quantified by the ratio with respect to the benchmark~$\text{SNR}^{*}_\text{coh}$, written as~\cite{Brambilla2008}
\begin{align}
\mathcal{R}_{s}=\frac{\text{SNR}^{*}_{s}}{\text{SNR}^{*}_{\text{coh}}}=\sqrt{\frac{2-\alpha}{\alpha^2Q+2(1-\alpha)\sigma+\alpha}},
\label{eq:Rstate}
\end{align}
where the subindex~$s$ denotes a state of light to be considered. The quantum gain is thus identified by~$\mathcal{R}_s >1$, meaning that a larger~$\text{SNR}^{*}$ is obtained in comparison with the case using a coherent state input. The ratio~$\mathcal{R}_{s}$ of Eq.~\eqref{eq:Rstate} can be reduced to~$\mathcal{R}_{s}\simeq\sigma^{-1/2}$ when~$\alpha\rightarrow 0$, and~$\mathcal{R}_{s}\simeq(Q+1)^{-1/2}$ when~$\alpha\rightarrow 0$. This implies that for highly transparent samples ($\alpha\simeq 0$), the quantum gain is dominantly obtained from the non-classical correlation, whereas for highly absorbing samples ($\alpha\simeq 1$), the quantum gain is dominantly obtained from the non-classical photon number statistics at the individual modes. It explains the reason why the TMSV state case becomes similar to the TF state case when~$\alpha\rightarrow0$ [see Fig.~\ref{fig:fig4}(b)] and to the coherent state case as~$\alpha\rightarrow1$ [see Fig.~\ref{fig:fig4}(c)], respectively. Also note that the quantum gain quantified by~${\cal R}$ is independent of~$\delta\alpha$ and~$\bar{c}$.

\section*{Effects of system imperfections}

To see how the ideal analysis shown above is affected by experimental imperfections from a practical point of view, we take into account photon loss with a rate~$\gamma$ and dark count contribution in both modes. The photon loss addresses not only scattering and absorption that occurs in optical components or transmission channels but also imperfect detection of the detectors considered in this work. Quantum gain of Eq.~\eqref{eq:Rstate} is significantly degraded by photon loss occurring via a probabilistic random process, which finally modifies the photon number distribution and breaks the quantum correlations between the two modes. As for the background, we considered the external stray light and the internal spontaneous photocurrent of the detector which is often called dark count. The background counts~$\{ n_\text{dark} \}$ are assumed to follow Poissonian distribution since the internal photocurrent takes place randomly~\cite{Kang2003, Excelitas_2020}, so reasonably assume that~$\langle (\Delta \hat{n}_\text{dark})^2\rangle=\mean{\hat{n}_\text{dark}}=N_\text{d}$. In the presence of photon loss with~$\gamma$, the transmission coefficients~$t_1$ and~$t_2$ can be expressed as~$t_1^2=(1-\alpha)(1-\gamma)$ and~$t_2^2=1-\gamma$, respectively. Including the dark count, one can write the photon number variances of the two modes as 
\begin{align*}
\langle(\Delta \hat{n}'_1)^2\rangle_\text{out}&=(1-\alpha)^2(1-\gamma)^2\langle(\Delta\hat{n}_1)^2\rangle_\text{in}+\{(1-\alpha)(1-\gamma)(\alpha+\gamma-\alpha\gamma)+\eta\}N,\nonumber\\
\langle(\Delta \hat{n}'_2)^2\rangle_\text{out}&=(1-\gamma)^2\langle(\Delta\hat{n}_1)^2\rangle_\text{in}+\{\gamma(1-\gamma)+\eta\}N,
\end{align*}
where~$\eta=N_\text{d}/N$ represents the ratio of the mean background count to the mean photon number of the input state. Then, the signal and noise modified by the imperfections are
\begin{align}
\Gamma_{-}^\text{imp}&= \delta\alpha(1-\gamma)N=(1-\gamma)\Gamma_{-},\nonumber\\
(\Delta I_{-}^\text{imp})^2&= (1-\gamma)^2\left[(\Delta I_{-})^2+\frac{\gamma(2-\alpha)N}{1-\gamma}+\frac{2\eta N}{(1-\gamma)^2}\right],\nonumber
\end{align}
respectively. Therefore, the~$\text{SNR}^{*}$ for~$\Gamma_-^\text{imp}$ can be described as
\begin{align}
\text{SNR}^{*}(\Gamma_{-}^\text{imp})=\frac{\sqrt{N_\text{eff}}}{\sqrt{\alpha^2Q+2(1-\alpha)\sigma+\alpha+\frac{\gamma(2-\alpha)}{1-\gamma}+\frac{2\eta}{(1-\gamma)^2}}}.
\label{eq:snr2}
\end{align}
Before looking into the effect of system imperfection in~$\text{SNR}^{*}$ of Eq.~\eqref{eq:snr2}, let us discuss the typical ranges of the respective imperfection parameters, which are found under normal experimental conditions. First, the detection efficiency of SPCMs is in a range between 0.2 and 0.95, including the recently developed superconducting nanowire single-photon detectors \cite{Eisaman2011, Semenov2001, Slussarenko2017}. Photon loss occurring in optical components and in transmission channels, on the other hand, varies depending on detailed experimental conditions. Hence, we consider the loss rate~$\gamma$ in a range from 0 to 1 in this work for generality. Note nevertheless that moderate values of~$\gamma$ only make sense because experiments with very small~$\gamma$ are not realistic and experiments with too large~$\gamma$ should be avoided. Second, and at the same time, currently available SPCMs suffer from the dead time ($\sim10^{-7}$s), for which the detector is irresponsive to any arriving photons. 
The dark count events, on the other hand, occur with~$10^1\sim10^3$ cps, implying that the value of~$\eta$ can thus be set in a range~$10^{-6}\sim10^{-1}$ when the input flux is assumed to be~$10^4\sim10^7$ cps.

\begin{figure}
\centering
\includegraphics[width=0.48\textwidth]{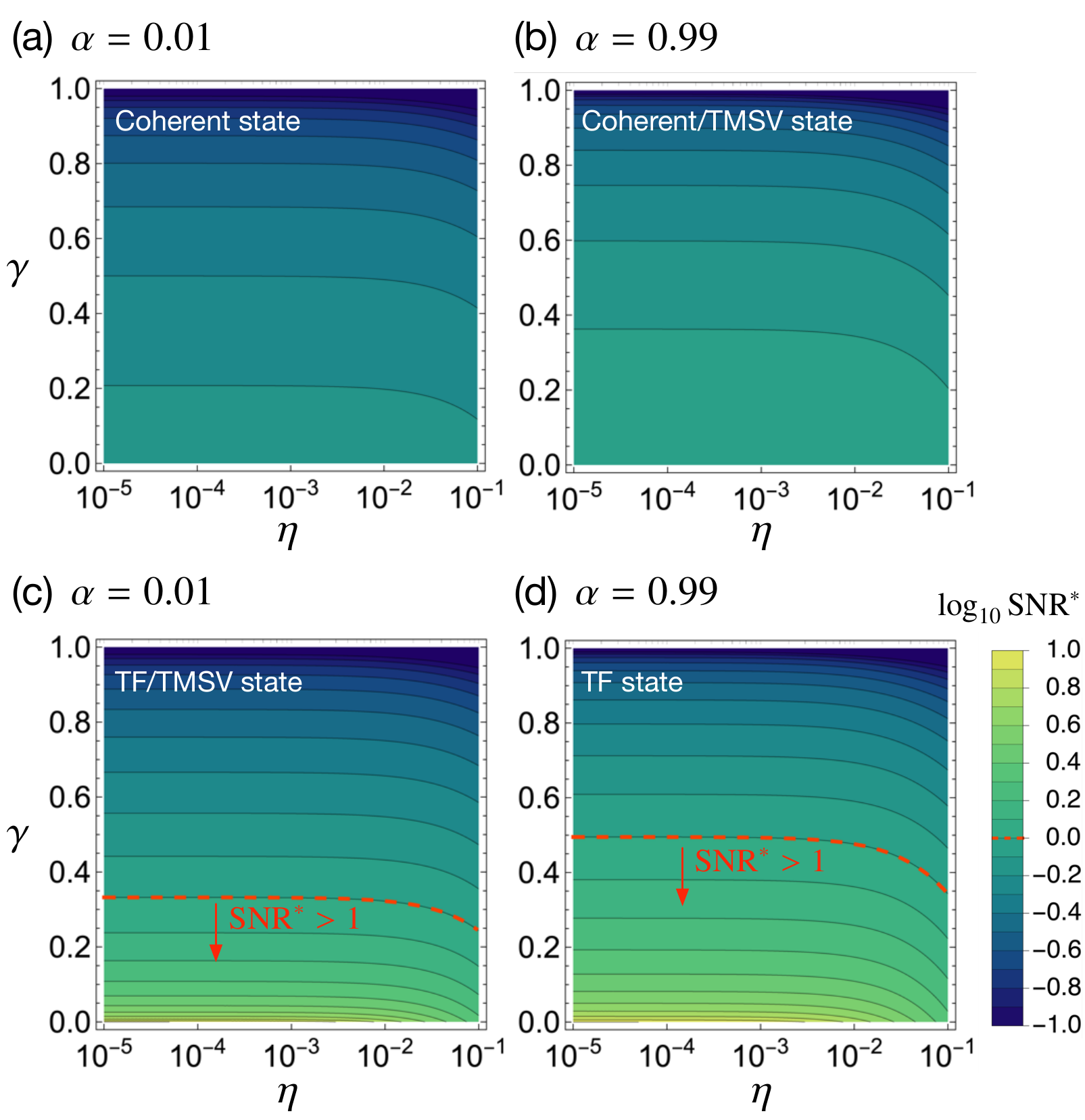}
\caption{$\text{SNR}^{*}$ of Eq.~\eqref{eq:snr2} in logarithmic scale as functions of~$\gamma$ and~$\eta$ for~$N_\text{eff}=1$ assumed as an example. The~$\text{SNR}^{*}$s for the cases using the coherent (upper panel), TF state (low panel), and TMSV state (upper and lower panels where appropriate) are respectively shown in the regimes of small~$\alpha$ (left column) and large~$\alpha$ (right column). In lower panel, the region where~$\text{SNR}^{*}>1$ is clearly distinguished by the red dashed line according to Eq.~\eqref{eq:gcr}.}
\label{fig:fig5}
\end{figure}

From now on, let us concentrate on two particular regimes of interest: small~$\alpha$ close to zero (i.e., highly transparent analytes) and large~$\alpha$ close to unity (i.e., highly absorbing analytes), where a considerable quantum enhancement in~$\text{SNR}^{*}$ has been observed in the absence of imperfections in the previous section. Figure~\ref{fig:fig5} shows~$\text{SNR}^{*}$s as functions of~$\gamma$ and~$\eta$ in the two regimes of~$\alpha=0.01$ and~$\alpha=0.99$ for $N_\text{eff}$ chosen as an example corresponding to the case that~$\delta\alpha=10^{-3}$ and~$\bar{c}=10^6$.  Figures~\ref{fig:fig5}(a) and (b) present the~$\text{SNR}^{*}$ for the case using a product coherent state input, setting the classical benchmark. Note that the~$\text{SNR}^{*}$ is always smaller than unity for both cases of small and large~$\alpha$, but can be improved by increasing~$N_\text{eff}$. Figures~\ref{fig:fig5}(c) and (d) show the~$\text{SNR}^{*}$ for the case using the TF state input. For both cases of small and large~$\alpha$, one can find a particular regime of~$\gamma$ and~$\eta$, where the~$\text{SNR}^{*}$ greater than unity can be achieved (see regions indicated by dashed line). The~$\text{SNR}^{*}$ for the TSMV state input coincides with that for the TF state input in the limit of small~$\alpha$ [see Fig.~\ref{fig:fig5}(c)] and with that for the product coherent state input in the limit of large~$\alpha$ [see Fig.~\ref{fig:fig5}(b)], as discussed before. It is evident to see that the~$\text{SNR}^{*}$ is robust to the background~$\eta$, but becomes degraded conspicuously when~$\eta>0.01$. The effect of the background with~$\eta$ can be understood in the critical loss rate~$\gamma_\text{c}$ defined as the maximum acceptable loss rate for~$\text{SNR}^{*}>1$, which is generally written as
\begin{align}
\gamma_\text{c}&=1-\frac{2-\alpha}{2\mathcal{N}}\left(1+\sqrt{1+\frac{8\eta\mathcal{N}}{(2-\alpha)^2}}\right)\simeq 1-\frac{2-\alpha}{\mathcal{N}}-\frac{2\eta}{2-\alpha},\text{  (for small~$\eta$)}
\label{eq:gcr}
\end{align}
where~$\mathcal{N}=N_\text{eff}-\alpha^2Q+2(1-\alpha)(1-\sigma)$. This clearly shows that the background effect becomes significant when the third term in Eq.~\eqref{eq:gcr} is not negligible, e.g., when~$\eta$ is greater than about 0.01. For negligible contribution of the background ($\eta\ll1$),~$\gamma_\text{c}\simeq 1-(2-\alpha)/\mathcal{N}$ and~$\text{SNR}^{*}>1$ can be achieved with the TF state input when~$\gamma<0.332$ and~$\gamma<0.495$ for the cases of~$\alpha=0.01$ and~$\alpha=0.99$, respectively. Although the~$\text{SNR}^{*}$ depends sensitively on the photon loss and the background, the quantum gain when using the TF state or TMSV state is always greater than unity in the whole range of the system parameters.

\begin{figure}
\centering
\includegraphics[width=0.48\textwidth]{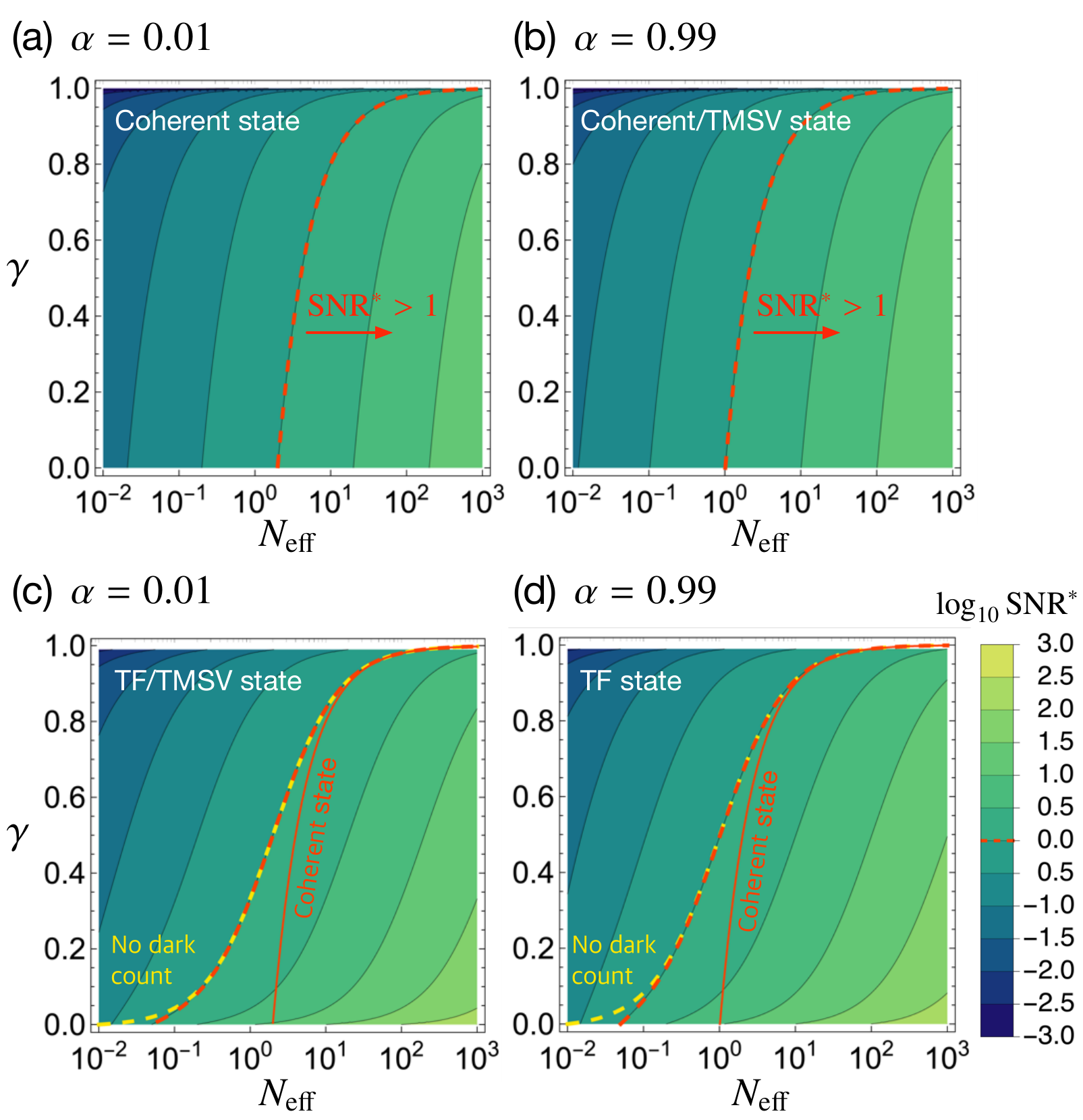}
\caption{
$\text{SNR}^{*}$ of Eq.~\eqref{eq:snr2} in logarithmic scale as functions of~$\gamma$ and~$N_\text{eff}$ for~$\mean{\hat{n}_\text{dark}}=10^3$ and~$\delta\alpha=10^{-3}$ assumed as an example. The~$\text{SNR}^{*}$s for the cases using the coherent (upper panel), TF state (low panel), and TMSV state (upper and lower panels where appropriate) are respectively shown in the regimes of small alpha (left column) and large alpha (right column). The region where~$\text{SNR}^{*}>1$ is clearly distinguished in both upper and lower panel by the red dashed line according to Eq.~\eqref{eq:gcr}. In lower panel, solid lines denote the boundary of~$\text{SNR}_\text{coh}^{*}=1$ for the coherent state input as in the upper panel, distinguishing the region with no quantum gain. Yellow dashed lines refers to the case of~$\mean{\hat{n}_\text{dark}}=0$.
}
\label{fig:fig6}
\end{figure}

As already noticed above, the effective mean photon number~$N_\text{eff}$ needs to be large enough to achieve~$\text{SNR}^{*}>1$ for given system parameters. To see the role of~$N_\text{eff}$ in details, let us now elaborate on the~$\text{SNR}^{*}$ as a function of the effective photon number~$N_\text{eff}$ and the loss rate~$\gamma$ for the four cases considered in Fig.~\ref{fig:fig5}. Here, we assume~$N_\text{d}=10^3$ as an example. Figure~\ref{fig:fig6} shows that the required~$N_\text{eff}$ for~$\text{SNR}^{*}>1$ increases as the system loss increases. It is also clear that the TF state requires smaller~$N_\text{eff}$ for~$\text{SNR}^{*}>1$ in comparison with the case using the coherent state input [see solid and dashed lines in Figs.~\ref{fig:fig6}(c) and (d)]. The effect of the background noise in the required effective photon number is only noticeable when system loss is very small [see dashed lines marked by~$\eta=0$ in Figs.~\ref{fig:fig6}(c) and (d)], so it is negligible in most cases except highly lossy conditions.

The aforementioned behaviors can also be understood analytically. In the individual regimes of small~$\alpha$ (i.e., highly transparent samples) and large~$\alpha$ (i.e., highly absorbing samples), one can derive from Eq.~\eqref{eq:snr2} the conditions for~$\text{SNR}^{*}>1$, written as
\begin{align}
\text{Small~$\alpha$: }&N_\text{eff}>2\sigma+\frac{2\gamma}{1-\gamma}+\frac{2\eta}{(1-\gamma)^2},
\label{eq:Neff1}
\end{align}
\begin{align}
\text{Large~$\alpha$: }&N_\text{eff}>Q+\frac{1}{1-\gamma}+\frac{2\eta}{(1-\gamma)^2}.
\label{eq:Neff2}
\end{align}
Note that the last two terms in both regimes are common for all twin-mode states under study in this work, but the first terms play a crucial role in determining the minimum~$N_\text{eff}$ for~$\text{SNR}^{*}>1$. The first terms depend on the kind of a twin-mode state input. In the first regime (small $\alpha$), the condition of Eq.~\eqref{eq:Neff1} has no dependence of~$Q$, so the TMSV state input exhibits the same behavior as the TF state input. In the second regime (large $\alpha$), on the other hand, the condition of Eq.~\eqref{eq:Neff2} is independent of~$\sigma$, the TMSV state input behaves as the product coherent state. For large values of~$\gamma$ (i.e., highly lossy) and~$\eta$ (i.e., highly noisy), the last two terms in Eqs.~\eqref{eq:Neff1} and~\eqref{eq:Neff2} predominate over the first terms with~$\sigma$ and~$Q$, respectively, so the conditions of Eqs.~\eqref{eq:Neff1} and~\eqref{eq:Neff2} becomes independent of the kind of a twin-mode state input. In other words, all the twin-mode states yield nearly the same~$\text{SNR}^{*}$ when setup is extremely lossy and noisy, consequently causing no quantum enhancement over the product coherent state. 

The quantum gains obtainable from the use of TF and TMSV states have already been shown clearly in Fig.~\ref{fig:fig4}(b), but it would be also meaningful to discuss the required brightness of a probe for~$\text{SNR}^{*}>1$ since using less intense input resources is more favorable to minimize the optical damages on an analyte. As discussed above, the effect of the background noise is negligible unless system loss is too small. Putting~$\eta=0$ in good approximation, one can simplify the conditions of Eqs.~\eqref{eq:Neff1} and~\eqref{eq:Neff2} for~$\text{SNR}^{*}>1$, which are shown in Table~\ref{tab:table1} for the three states at the two regimes of interest.
\begin{table}[ht]
\centering
\begin{tabular}{|c|c|c|c|}
\hline
 & Coherent state & TMSV state &  TF state \\
\hline 
Small~$\alpha$ &~$N_\text{eff}>\frac{2}{1-\gamma}$ &~$N_\text{eff}>\frac{2\gamma}{1-\gamma}$ &~$N_\text{eff}>\frac{2\gamma}{1-\gamma}$\\
\hline
Large~$\alpha$ &~$N_\text{eff}>\frac{1}{1-\gamma}$ &~$N_\text{eff}>\frac{1}{1-\gamma}$
&~$N_\text{eff}>\frac{\gamma}{1-\gamma}$\\
\hline
\end{tabular}
\caption{Conditions of~$N_\text{eff}$ such that~$\text{SNR}>1$ for the coherent, TF and TMSV state inputs, assuming~$\eta=0$ in good approximation due to its negligible contribution for moderate $\gamma$.}
\label{tab:table1} 
\end{table}

\section*{Conclusion}
We have studied the use of practical quantum state inputs in quantum imaging in comparison with the case using a coherent state input. As an imaging scheme, we have employed a differential measurement scheme based on time-integrated single-photon detection. The defect-detection capability $\text{SNR}^{*}$s have been investigated in detail with and without experimental imperfections, respectively. We have shown that the $\text{SNR}^{*}$s is significantly degraded by the system loss, but the quantum gain over the classical benchmark is still achievable under a moderate background noise. We have also discussed quantitatively when the background noise starts to significantly affect the $\text{SNR}^{*}$s. Our analyses clearly demonstrate what quantum characteristic of quantum state inputs play a critical role for a given experimental environment. Since the two parameters of Mandel Q-parameter and the noise reduction factor $\sigma$ of light sources are used as the characteristics of used light sources, any other quantum state can be easily estimated its sensing performance by simply investigating those two parameter values for the balanced detection scheme. With the analyses provided in this work, we hope that our result can be used as a guide for designing practical quantum optical imaging schemes of an absorptive analyte under realistic conditions.
An interesting scenario found in our analyses is that, even for a highly absorptive analyte, the $\text{SNR}^{*}$ can diverge when the TF state is utilized as an input. This will be useful for monitoring a small defect embedded in an opaque sample. Alternatively to the TF state input, we can consider an ideal single photon source which can be generated from a single emitter excited in a regular excitation interval. Within current technology, the single emitters can be used to produce partially indistinguishable single photons with good quality~\cite{Aharonovich2016,Chu2017,Wang2019}. The purity and the fidelity of those single photons at room temperature are close to the unity, although several obstacles are still to be resolved. It would thus be interesting to implement those single photons for detecting small detects in opaque conditions.

\section*{Acknowledgments}
This study was supported by the Basic Science Research Program through the National Research Foundation (NRF) of Korea and funded by the Ministry of Science and ICT (No. 2020R1A2C1010014) and Institute of Information and Communications Technology Planning and Evaluation (IITP) grant funded by the Korea government (MSIT) (No. 2019-0-00296), and by Korea Research Institute of Standards and Science (KRISS–GP2022-0012).

\section*{Supplementary}
In this supplementary material, we show that the $\text{SNR}^{*}$ of Eq.~(\ref{eq:SNR_star}) calculated for PNRD is approximately equal to the $\text{SNR}^{*}$ obtainable by the SPCM.

\subsection*{$\text{SNR}^*$ for click statistics}\label{appendix:initial_setup}
\setcounter{equation}{0}
\renewcommand{\theequation}{A\arabic{equation}}
\setcounter{figure}{0}
\renewcommand{\thefigure}{A\arabic{figure}}

When two local SPCMs are performed~$m$ times onto a two-mode state~$\rho$, they produce two sequential outcomes with a length of~$m$ for the individual modes, and the outcomes are recorded in terms of~$0$ (i.e., `no click') and~$1$ (i.e., `click'). The number of click events is then counted in each mode, so that we consider two-mode click statistics~$\{ c_1, c_2\}$ for the numbers of click events~$c_j\in \{ 0,1,\cdots m\}$. The statistical features of the click statistics need to be implemented in the calculation of~$\text{SNR}$ and~$\text{SNR}^{*}$ for a given state~$\rho$.

To this end, let us first consider the probability of the counts~$\boldsymbol{\nu}=(\nu_{00},\nu_{01},\nu_{10},\nu_{11})$ for the set of individual detection outcomes~$\left\{ 00,01,10,11 \right\}$ out of~$m$ sequential measurements. The subscript denotes the "no click" event for $0$ and "click" event for $1$, each for signal channel at the first place, and reference channel at the second place of the subscript. It follows the multinomial distribution written as
\begin{equation*}
P(\boldsymbol{\nu}) 
= \frac{m!}{\nu_{00}!\nu_{01}!\nu_{10}!\nu_{11}!}{P_{00}}^{\nu_{00}}{P_{01}}^{\nu_{01}}{P_{10}}^{\nu_{10}}{P_{11}}^{\nu_{11}},
\end{equation*}
where~$\left\vert\!\left\vert \boldsymbol{\nu} \right\vert\!\right\vert_1=m$, ~$P_{\mu}=\text{Tr}[\hat{\pi}_{\mu}\rho]$ represents the probability for the outcome~$\mu$ at individual measurements, and~$\hat{\pi}_{\mu}=\vert \mu \rangle \langle \mu \vert$ denotes the positive operator-valued measure (POVM) projector corresponding to the outcome~$\mu\in\left\{ 00,01,10,11 \right\}$.
Then, the expectation value, variance, and covariance for the counts of individual events are written as
\begin{align*}
\text{E}(\nu_\mu) &= m P_{\mu},\\
\text{Var}( \nu_{\mu} ) &= mP_{\mu}( 1 - P_{\mu} ),\\
\text{Cov}( \nu_{\mu},\nu_{\beta} ) &= - mP_{\mu}P_{\beta},
\end{align*}
respectively. Therefore, the expectation value and the variance of the click counts~$c_j$ for each mode can be written as
\begin{align*}
\mean{c_1}&=\text{E}(\nu_{10})+\text{E}(\nu_{11})=m(P_{10}+P_{11}),\\
\mean{c_2}&=\text{E}(\nu_{01})+\text{E}(\nu_{11})=m(P_{01}+P_{11}),\\
\mean{\Delta^2 c_1}&=\text{Var}( \nu_{10} )+\text{Var}( \nu_{11} )+2\text{Cov}( \nu_{10},\nu_{11} )
=m\left[P_{10}+P_{11}-(P_{10}+P_{11})^2\right],\\
\mean{\Delta^2 c_2}&=\text{Var}( \nu_{01} )+\text{Var}( \nu_{11} )+2\text{Cov}( \nu_{01},\nu_{11} )
=m\left[P_{01}+P_{11}-(P_{01}+P_{11})^2\right].
\end{align*}
Here, it is obvious that the click statistics can be calculated for any state~$\rho$, but is different from the true photon number statistics that is obtainable by PNRD. For the differential imaging scheme, the expectation value and the variance of the differential click numbers between~$c_1$ and~$c_2$ are then given as
\begin{align*}
I_{-}&=\mean{c_{-}}=\text{E}(\nu_{01})-\text{E}(\nu_{10})=m(P_{01}-P_{10}),\\
\Gamma_{-}&=I_{-}(\alpha+\delta\alpha)-I_{-}(\alpha)=m(\delta P_{01}-\delta P_{10}),\\
(\Delta I_{-})^2 &=\mean{(\Delta c_{-})^2}=\text{Var}( \nu_{01} )+\text{Var}( \nu_{10} )-2\text{Cov}( \nu_{10},\nu_{01} )
=m\left[P_{10}+P_{01}-(P_{10}-P_{01})^2\right],
\end{align*}
where~$\delta P_{\mu}=P_{\mu}(\alpha+\delta\alpha)-P_{\mu}(\alpha)$ and~$P_{\mu}(\alpha)=\text{Tr}[\hat{\pi}_\mu \rho(\alpha)]$ for a state~$\rho$ having undergone absorption with~$\alpha$. The~$\text{SNR}^{*}$ is then
\begin{align}
\text{SNR}^{*}=\frac{\Gamma_{-}}{\Delta I_{-}}=\frac{\sqrt{m}(\delta P_{01}-\delta P_{10})}{\sqrt{P_{10}+P_{01}-(P_{10}-P_{01})^2}}.\label{append:eq:SNR*}
\end{align}
The calculation of~$\text{SNR}^*$ is now straightforward for any state~$\rho$. Normally the effect of detection inefficiency and dark counts are included in the POVM elements via proper modification of a detector model, but here we include them into a quantum state~$\rho$ for convenience, as further discussed below.

\subsection*{Lossy and noisy quantum states with~$N\ll1$}\label{appendix:system_imperfection}
\setcounter{equation}{0}
\renewcommand{\theequation}{B\arabic{equation}}
\setcounter{figure}{0}
\renewcommand{\thefigure}{B\arabic{figure}}

For the calculation of~$\text{SNR}^*$ of Eq.~\eqref{append:eq:SNR*}, here we include the effect of loss, absorption, and dark count into a two-mode quantum state~$\rho$. Let us begin with assuming equal loss rates~$\gamma_1=\gamma_2=\gamma$, equal mean dark counts~$\mean{\hat{n}_{\text{d}1}}=\mean{\hat{n}_{\text{d}2}}=N_{\text{d}}$, and a twin-mode input state, i.e.,~$\mean{\hat{n}_{1}}=\mean{\hat{n}_{2}}=N$, as in the main text for simplicity. We further assume that the contribution from higher photon numbers than a single photon is negligible, i.e.,~$N\ll 1$. In the limit of small~$N$, we can write an arbitrary twin-mode quantum state as
\begin{align}
\vert \Psi\rangle \approx \sum_{j,k=0}^{1}c_{j,k}\vert j,k\rangle,\label{eq:approx_state}
\end{align}
for which~$\sum_{j,k=0}^{1} \vert c_{j,k}\vert^2 \approx 1$ and~$\langle \Psi\vert \hat{n}_{1,2}\vert \Psi\rangle \approx N \ll 1$. Since SPCM (and even PNRD) access only the diagonal elements of a density matrix~$\vert \Psi\rangle\langle \Psi\vert$, an effective density matrix for the state of Eq.~\eqref{eq:approx_state} can be written in an operational sense as
\begin{align}
\rho_\text{in} \simeq p_{00}\hat{\pi}_{00} + p_{01}\hat{\pi}_{01} + p_{10}\hat{\pi}_{10} + p_{11}\hat{\pi}_{11},\label{eq:approx_matrix_ideal}
\end{align}
where~$p_{jk}=\vert c_{jk}\vert^2$ denotes the population of the state $\vert j,k\rangle$ in $\rho$ and here the projectors~$\hat{\pi}_{jk}$ are used as the photon-number basis of a density matrix. Here, $p_{jk}$ are determined for a given input state.

The absorption ($\alpha$) and loss ($\gamma$) can be implemented by modifying $p_{jk} \rightarrow p_{jk}(\alpha,\gamma)$ in the ideal state of Eq.~\eqref{eq:approx_matrix_ideal}, where 
\begin{align*}
p_{00}(\alpha,\gamma)&=p_{00}+\gamma p_{01}
+[\alpha+\gamma(1-\alpha)]p_{10}+\gamma[\alpha+\gamma(1-\alpha)]p_{11},\\
p_{01}(\alpha,\gamma)&=(1-\gamma)(p_{01}+[\alpha+\gamma(1-\alpha)]p_{11}),\\
p_{10}(\alpha,\gamma)&=(1-\alpha)(1-\gamma)(p_{10}+\gamma p_{11}),\\
p_{11}(\alpha,\gamma)&=(1-\alpha)(1-\gamma)^{2}p_{11},
\end{align*}
leading to
\begin{align}
\rho_\text{lossy}
\simeq p_{00}(\alpha,\gamma)\hat{\pi}_{00} +p_{01}(\alpha,\gamma)\hat{\pi}_{01}
+p_{10}(\alpha,\gamma)\hat{\pi}_{10} 
+p_{11}(\alpha,\gamma)\hat{\pi}_{11}.
\label{eq:approx_matrix_lossy}
\end{align}

The dark count can be modeled by an inflow of a weak coherent state in addition to the actual states being measured, so it can be implemented by modifying $\hat{\pi}_{jk}\rightarrow\hat{\pi}_{jk}'$ in the lossy state of Eq.~\eqref{eq:approx_matrix_lossy}, where
\begin{align*}
\hat{\pi}_{00}'&=e^{-2N_\text{d}}\hat{\pi}_{00}+e^{-N_\text{d}}(1-e^{-N_\text{d}})(\hat{\pi}_{01}+\hat{\pi}_{10})+(1-e^{-N_\text{d}})^{2}\hat{\pi}_{11},\\
\hat{\pi}_{01}'&=e^{-N_\text{d}}\hat{\pi}_{01}+(1-e^{-N_\text{d}})\hat{\pi}_{11},\\
\hat{\pi}_{10}'&=e^{-N_\text{d}}\hat{\pi}_{10}+(1-e^{-N_\text{d}})\hat{\pi}_{11},\\
\hat{\pi}_{11}'&=\hat{\pi}_{11}.
\end{align*}
Here, $N_\text{d}$ is the average inflow photon number. Therefore, a lossy and noisy state through the modification introduced above can be written as
\begin{align}
\rho_\text{lossy\&noisy}\simeq p_{00}(\alpha,\gamma)\hat{\pi}_{00}' + p_{01}(\alpha,\gamma)\hat{\pi}_{01}' + p_{10}(\alpha,\gamma)\hat{\pi}_{10}' + p_{11}(\alpha,\gamma)\hat{\pi}_{11}'.
\label{eq:approx_matrix_lossy_noisy}
\end{align}

\subsection*{Investigation of various input states}\label{appendix:source_characterization}
\setcounter{equation}{0}
\renewcommand{\theequation}{C\arabic{equation}}
\setcounter{figure}{0}
\renewcommand{\thefigure}{S\arabic{figure}}


The state of Eq.~\eqref{eq:approx_matrix_lossy_noisy} is of the general form for an arbitrary two-mode state when $N\ll1$. Here we investigate the $\text{SNR}^*$'s of Eq.~\eqref{append:eq:SNR*} for various input states and compare those with ~$\text{SNR}^*$ for the true photon number statistics. For the calculation of $\text{SNR}^*$'s with respect to a state $\rho$ of Eq.~\eqref{eq:approx_matrix_lossy_noisy}, one can write $P_\mu=\text{Tr}[\hat{\pi}_{\mu}\rho]$ for $\mu\in\{00,01,10,11\}$:
\begin{align}
P_{00}(\alpha,\gamma)&=e^{-2N_\text{d}}p_{00}(\alpha,\gamma),\label{eq:P00_final}\\
P_{01}(\alpha,\gamma)&=e^{-N_\text{d}}p_{01}(\alpha,\gamma)+e^{-N_\text{d}}(1-e^{-N_\text{d}})p_{00}(\alpha,\gamma),\label{eq:P01_final}\\
P_{10}(\alpha,\gamma)&=e^{-N_\text{d}}p_{10}(\alpha,\gamma)+e^{-N_\text{d}}(1-e^{-N_\text{d}})p_{00}(\alpha,\gamma),\label{eq:P10_final}\\
P_{11}(\alpha,\gamma)&=p_{11}(\alpha,\gamma)+(1-e^{-N_\text{d}})\left[p_{01}(\alpha,\gamma)+p_{10}(\alpha,\gamma)\right]+(1-e^{-N_\text{d}})^{2}p_{00}(\alpha,\gamma),\label{eq:P11_final}
\end{align}
where $p_{\mu}(\alpha, \gamma)$ are to be set by a given input state. Below, writing $p_{\mu}(\alpha,\gamma)$ for the three states considered in the main text, we obtain $P_{\mu}(\alpha,\gamma)$ using Eqs.~\eqref{eq:P00_final}-\eqref{eq:P11_final}, leading to the calculation of $\text{SNR}^*$'s.

\subsubsection*{Coherent state input}
The coherent state input can be written in the form of Eq.~\eqref{eq:approx_matrix_lossy_noisy} as $\rho^{\text{coh}}\simeq e^{-2N}\hat{\pi}_{00}+e^{-N}(1-e^{-N})(\hat{\pi}_{01}+\hat{\pi}_{10})+(1-e^{-N})^{2}\hat{\pi}_{11}$, i.e., 
\begin{align*}
p_{00}&=e^{-2N},\\
p_{01}&=p_{10}=e^{-N}(1-e^{-N}),\\
p_{11}&=(1-e^{-N})^{2},
\end{align*}
leading to
\begin{align*}
p_{00}(\alpha,\gamma)&=e^{-2N}+[\alpha+\gamma(2-\alpha)]e^{-N}(1-e^{-N})+\gamma(\alpha+\gamma-\alpha\gamma)(1-e^{-N})^{2},\\
p_{01}(\alpha,\gamma)&=(1-\gamma)(1-e^{-N})[e^{-N}+(\alpha+\gamma-\alpha\gamma)(1-e^{-N})],\\
p_{10}(\alpha,\gamma)&=(1-\alpha)(1-\gamma)(1-e^{-N})[e^{-N}+\gamma (1-e^{-N})],\\
p_{11}(\alpha,\gamma)&=(1-\alpha)(1-\gamma)^{2}(1-e^{-N})^{2}.
\end{align*}
Therefore, $P_{\mu}$'s are written as
\begin{align*}
P_{00}(\alpha,\gamma)&=e^{-2N_{\text{d}}}\{e^{-N}+\gamma(1-e^{-N})\}\left[e^{-N}+(\alpha+\gamma-\alpha\gamma)(1-e^{-N})\right],\\
P_{01}(\alpha,\gamma)&=e^{-N_{\text{d}}}\{1-e^{-N-N_{\text{d}}}-\gamma e^{-N_{\text{d}}}(1-e^{-N})\}\left[e^{-N}+(\alpha+\gamma-\alpha\gamma)(1-e^{-N})\right],\\
P_{10}(\alpha,\gamma)&=e^{-N_{\text{d}}}\{e^{-N}+\gamma(1-e^{-N})\}\left[1-(\alpha+\gamma-\alpha\gamma)e^{-N_{\text{d}}}-(1-\alpha)(1-\gamma)e^{-N-N_{\text{d}}}\right],\\
P_{11}(\alpha,\gamma)&=\{1-e^{-N-N_{\text{d}}}-\gamma e^{-N_{\text{d}}}(1-e^{-N})\}\left[1-(\alpha+\gamma-\alpha\gamma)e^{-N_{\text{d}}}-(1-\alpha)(1-\gamma)e^{-N-N_{\text{d}}}\right].
\end{align*}
The signal and variance for coherent state are then
\begin{align*}
\Gamma_{-}^{\text{coh}}&=m\delta\alpha e^{-N_{\text{d}}}(1-\gamma)(1-e^{-N}),\\
(\Delta I_{-}^{\text{coh}})^{2}&=me^{-N_{\text{d}}}[\alpha(1-\gamma)+2\gamma+e^{-N}(2-\alpha)(1-\gamma)
-e^{-2N-N_{\text{d}}}(2-\alpha(2-\alpha))(1-\gamma)^{2}\\
&\quad-2e^{-N-N_{\text{d}}}(1-\gamma)(\alpha-\alpha^{2}(1-\gamma)+2\gamma-2\alpha\gamma)
-e^{-N_{\text{d}}}(\alpha^{2}(1-\gamma)^{2}+2\alpha\gamma(1-\gamma)+2\gamma^{2})].
\end{align*}

\subsubsection*{TMSV state input}
For a TMSV state, the photon numbers are completely correlated between the two modes, so it can be written in the form of Eq.~\eqref{eq:approx_matrix_lossy_noisy} as 
\begin{align*}
\rho^{\text{TMSV}}
\simeq p_{00}\hat{\pi}_{00}+p_{11}\hat{\pi}_{11},
\end{align*}
where 
$p_{00}=e^{-N}$ [i.e., $p_{01}=p_{10}=0$ and $p_{11}=(1-e^{-N})$], originated from the Poisson distribution due to the multi-mode feature of a TMSV state. 
This leads to
\begin{align*}
p_{00}(\alpha,\gamma)&=e^{-N}+\gamma(\alpha+\gamma-\alpha\gamma)(1-e^{-N}),\\
p_{01}(\alpha,\gamma)&=(1-\gamma)(\alpha+\gamma-\alpha\gamma)(1-e^{-N}),\\
p_{10}(\alpha,\gamma)&=(1-\alpha)\gamma(1-\gamma)(1-e^{-N}),\\
p_{11}(\alpha,\gamma)&=(1-\alpha)(1-\gamma)^{2}(1-e^{-N}).
\end{align*}
Therefore, $P_{\mu}$'s are written
\begin{align*}
P_{00}(\alpha,\gamma)&=e^{-2N_{\text{d}}}\left[e^{-N}+\gamma(\alpha+\gamma-\alpha\gamma)(1-e^{-N})\right],\\
P_{01}(\alpha,\gamma)&=e^{-N_{\text{d}}}\left[(\alpha+\gamma-\alpha\gamma)(1-\gamma e^{-N_\text{d}})-e^{-N}(1-\gamma)\{e^{-N_\text{d}}-(1-\alpha)(1-\gamma e^{-N_\text{d}})\}\right],\\
P_{10}(\alpha,\gamma)&=e^{-N_{\text{d}}}\left[\gamma\{1-(\alpha+\gamma-\alpha\gamma)e^{-N_\text{d}}\}+e^{-N}(1-\gamma)\{1-e^{-N_\text{d}}-\gamma(1-\alpha)e^{-N_\text{d}}\}\right],\\
P_{11}(\alpha,\gamma)&=(1-\gamma e^{-N_\text{d}})\{1-(\alpha+\gamma-\alpha\gamma)e^{-N_\text{d}}\}+e^{-N-N_{\text{d}}}(1-\gamma)\{e^{-N_\text{d}}-(2-\alpha)+\gamma(1-\alpha)e^{-N_\text{d}}\}.
\end{align*}
The signal and variance for the TMSV state input are then
\begin{align*}
\Gamma_{-}^{\text{TMSV}}&=m\delta\alpha e^{-N_{\text{d}}}(1-\gamma)(1-e^{-N})\\
(\Delta I_{-}^{\text{TMSV}})^{2}&=me^{-N_{\text{d}}}\left[2e^{-N}(1-e^{-N_{\text{d}}})+(1-e^{-N})\{\alpha(1-\gamma)+2\gamma
-e^{-N_{\text{d}}}\left(2\gamma(\alpha+\gamma-\alpha\gamma)+\alpha^{2}(1-e^{-N})(1-\gamma)^{2}\right)\}\right].
\end{align*}

\subsubsection*{TF state input}

\begin{figure}[h]
\centering
\includegraphics[width=0.5\textwidth]{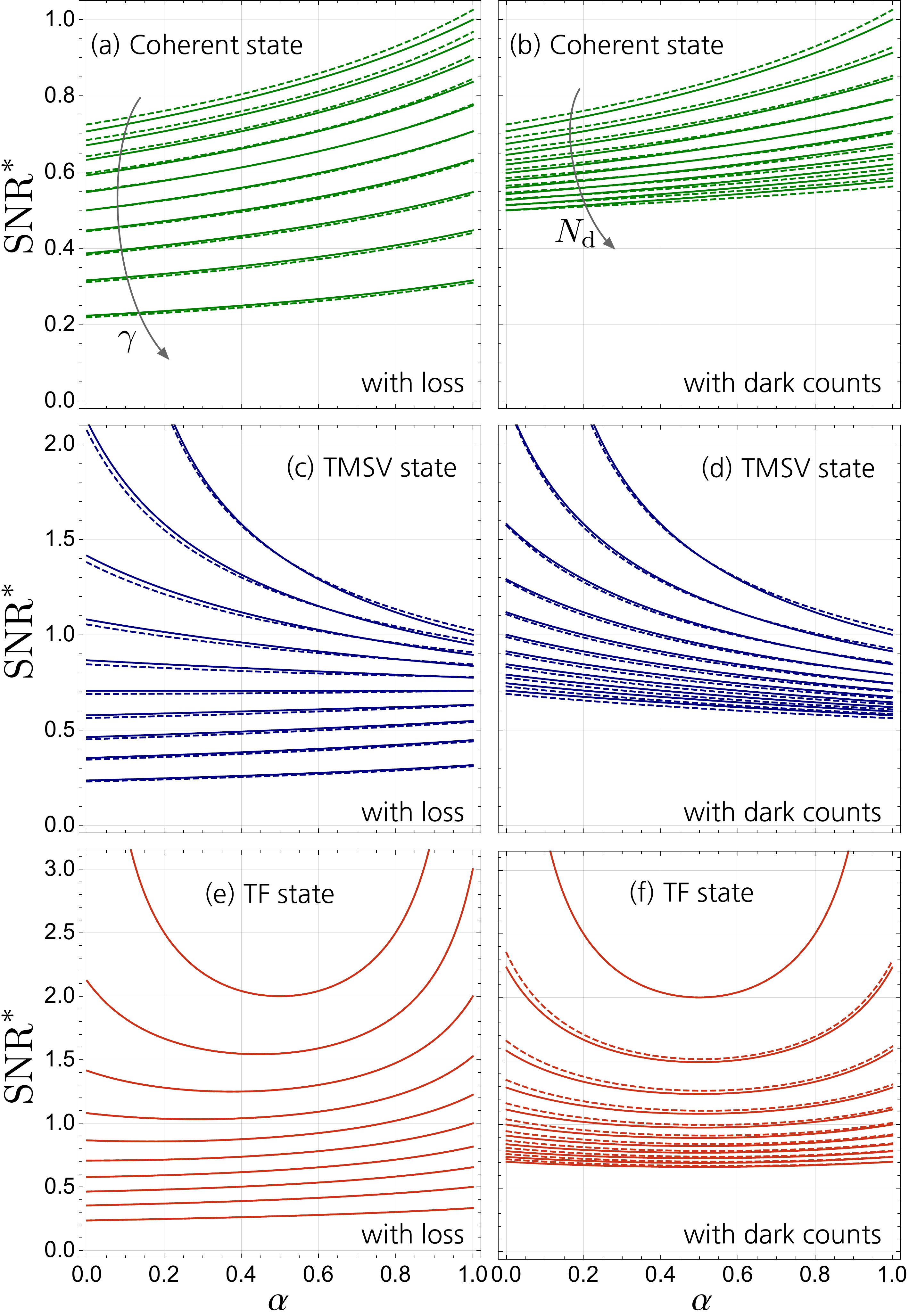}
\caption{
$\text{SNR}^{*}$ as a function of~$\alpha$ for various input states (respective rows) when using SPCMs (dashed curves) and PNRD (solid curves) with loss (left column) and the dark counts (right column). 
Here, we consider $\gamma=0,0.1,\cdots,0.9$ for loss and $N_\text{d}=0,0.01,\cdots,0.1$ for the dark counts, showing that $\text{SNR}^{*}$ for all the cases generally decreases with loss and dark counts. Particularly for SPCMs,~$m=10^{7}$ and~$\delta\alpha=10^{-3}$ have used.}
\label{fig:fig7}
\end{figure}

Although the state of Eq.~\eqref{eq:approx_matrix_lossy_noisy} can be used for an arbitrary two-mode state, we here consider the TF state input as the case that leads to $mN$ detection events among $m$ time bins with an assumption $mN \in \mathbb{Z}$ and $N\ll 1$. In other words, one can consider that a sequence of $mN$ TF states fires the click events at individual time bins among $m$ bins, but randomly distributed. Such an assumption is set for a reasonable comparison with the other states considered above. At time bins where the TF state is not given, the click event can also be fired due to the dark count contribution. 
The statistical features of the click number distribution are thus given as
\begin{align*}
E\left( \nu_{\mu}\right) &= mNP_{\mu}+m(1-N)P^{\text{dark}}_{\mu},\\
\text{Var}\left( \nu_{\mu} \right) &= mNP_{\mu}( 1 - P_{\mu} )+m(1-N)P^{\text{dark}}_{\mu}(1-P^{\text{dark}}_{\mu}),\\
\text{Cov}\left( \nu_{\mu},\nu_{\beta} \right) &= - mNP_{\mu}P_{\beta},
\end{align*}
where~$P_{\mu}=\text{Tr}(\rho\hat{\pi}_{\mu})$ for the TF state $\rho$ and $P^{\text{dark}}_{\mu}=\text{Tr}(\rho^{\text{dark}}\hat{\pi}_{\mu})$ for an effective coherent state $\rho^\text{dark}$ causing the dark count. Here, the second terms are due to dark count contribution for~$m(1-N)$ time bins and there is no correlation between the contributions from $\rho$ and $\rho^\text{dark}$. 

The SPCM cannot distinguish between a single photon and more photons, the TF state can be considered as $\rho^{\text{TF}}=\hat{\pi}_{11}$, i.e., a pair of single-photon states, for which 
\begin{align*}
p_{00}&=p_{01}=p_{10}=0,\\
p_{11}&=1,
\end{align*}
leading to
\begin{align*}
p_{00}(\alpha,\gamma)&=\gamma(\alpha+\gamma-\alpha\gamma),\\
p_{01}(\alpha,\gamma)&=(1-\gamma)(\alpha+\gamma-\alpha\gamma),\\
p_{10}(\alpha,\gamma)&=\gamma(1-\alpha)(1-\gamma),\\
p_{11}(\alpha,\gamma)&=(1-\alpha)(1-\gamma)^{2}.
\end{align*}
Therefore, $P_{\mu}$'s are written as
\begin{align*}
P_{00}(\alpha,\gamma)&=e^{-2N_{\text{d}}}\gamma(\alpha+\gamma-\alpha\gamma),\\
P_{01}(\alpha,\gamma)&=e^{-N_{\text{d}}}(1-\gamma e^{-N_{\text{d}}})(\alpha+\gamma-\alpha\gamma),\\
P_{10}(\alpha,\gamma)&=e^{-N_{\text{d}}}\gamma\left[1-(\alpha+\gamma-\alpha\gamma)e^{-N_{\text{d}}}\right],\\
P_{11}(\alpha,\gamma)&=(1-\gamma e^{-N_{\text{d}}})\left[1-(\alpha+\gamma-\alpha\gamma)e^{-N_{\text{d}}}\right].
\end{align*}

The signal and variance for the TF state input are then
\begin{align*}
\Gamma_{-}^{\text{TF}}&=mN\delta\alpha e^{-N_{\text{d}}}(1-\gamma),\\
(\Delta I_{-}^{\text{TF}})^2 
&=me^{-N_{\text{d}}}\left[2(1-e^{-N_{\text{d}}})-N(2-\alpha)(1-\gamma)-Ne^{-N_{\text{d}}}(1-\gamma)\{\alpha^{2}(1-\gamma)+2\alpha\gamma-2\gamma-2\}\right],
\end{align*}

\subsubsection*{$\text{SNR}^*$}

Figure~\ref{fig:fig7} presents the comparison of $\text{SNR}^{*}$ between the two detection schemes: SPCMs of Eq.~\eqref{append:eq:SNR*} and PNRD of Eq.~(5) for the three input states (see the respective rows) with loss (see the left panel) and with the dark count (see the right panel). For each input state, the cases using SPCMs and PNRD are represented by dashed and solid curves, respectively. It is clear to see that the two detection schemes are very similar in most cases. This indicates that the detection scheme using SPCMs for a weak input field (i.e., $N\ll 1$) can effectively realize the detection scheme using PNRD, which justifies the use of Eq.~(5) for the scheme using SPCMs in good approximation for simplicity and convenience.

\bibliography{reference.bib}

\end{document}